\documentclass[floatfix,twocolumn,showpacs,floats,aps,prl,showpacs,amsfonts,amssymb]{revtex4}
\usepackage{amssymb}
\usepackage{bm}
\usepackage{hyphenat}
\usepackage{graphicx}
\usepackage{psfrag}
\usepackage[latin1]{inputenc}
\newlength{\sepmod}
\def\eqref#1{(\ref{#1})}
\def\l{\langle}

\def\r{\rangle}

\begin{document}
\title{Ground States and Defect Energies of the Two-dimensional {\em XY\/} Spin Glass
from a Quasi-Exact Algorithm}

\author{Martin Weigel}
\altaffiliation{Present address: Department of Mathematics, Heriot-Watt University,
  Edinburgh, EH14~4AS, UK}

\author{Michel J.\ P.\ Gingras}
\affiliation{Department of Physics, University of Waterloo, Waterloo, Ontario,
  N2L~3G1, Canada}

\date{\today}

\begin{abstract}
  We employ a novel algorithm using a quasi-exact embedded-cluster matching technique
  as minimization method within a genetic algorithm to reliably obtain numerically
  exact ground states of the Edwards-Anderson {\em XY\/} spin glass model with
  bimodal coupling distribution for square lattices of up to $28\times 28$ spins.
  Contrary to previous conjectures, the ground state of each disorder replica is
  non-degenerate up to a global O(2) rotation. The scaling of spin and chiral defect
  energies induced by applying several different sets of boundary conditions exhibits
  strong crossover effects. This suggests that previous calculations have yielded
  results far from the asymptotic regime. The novel algorithm and the aspect-ratio
  scaling technique consistently give $\theta_s=-0.308(30)$ and $\theta_c=-0.114(16)$
  for the spin and chiral stiffness exponents, respectively.
\end{abstract}


\maketitle


Since the suggestion of Edwards and Anderson (EA) to capture the essence of spin
glass behavior in a class of simple lattice models thirty years ago
\cite{edwards:75a}, the quest for their understanding has spurred an enormous
research effort \cite{kawashima:03a}. EA considered the Hamiltonian
\begin{equation}
  \label{eq:EA_model}
  {\cal H} = -\sum_{\l ij\r}J_{ij}\,\bm{S}_i\cdot\bm{S}_j,
\end{equation}
with O($n$) spins $\bm{S}_i$ on a regular lattice with quenched, random and
frustrated nearest-neighbor interactions $J_{ij}$.  Although substantial progress has
been made in recent years in understanding Ising and vector spin glasses in finite
dimensions $D$, mostly by the development and application of sophisticated numerical
techniques, we still lack an undisputed theory of the spin glass phase
\cite{kawashima:03a}. Due to its relative simplicity, by far the most work has been
devoted to the Ising spin glass~\cite{kawashima:03a}. However, much less advance has
been made on models with continuous spins which are often more relevant to real
materials \cite{kawashima:03a}.

The properties of the spin glass phase in the EA model are described by a scaling
theory of the associated zero-temperature fixed point \cite{bray:87a}.  The
corresponding renormalization-group (RG) picture considers the scaling of the width
of the distribution of random couplings, $P_L(J_{ij})$, with the coarse-graining
length scale $L$, $J(L)\sim J L^{\theta_s}$, defining the spin stiffness exponent
$\theta_s$. Depending on whether $\theta_s>0$ or $\theta_s<0$, the spin glass phase
is stable or unstable against thermal fluctuations, respectively. Following a
suggestion by Banavar {\em et al.}\ and McMillan \cite{banavar:82a}, the scaling of
$J(L)$ can be inferred from monitoring the dependence of the energy of droplet or
domain-wall excitations induced by a change of boundary conditions (BCs), giving rise
to the name ``domain-wall RG'' (DWRG) method. For cases where $\theta_s<0$, and thus
the spin-glass transition temperature $T_g=0$, such as for the EA Ising model in two
dimensions (2D) \cite{kawashima:03a}, $\theta_s$ also determines the critical
behavior with the spin glass correlation length diverging as $\xi\sim T^{-\nu_s}$ for
$T\downarrow 0$, where $\nu_s=-1/{\theta_s}$ \cite{bray:87a}. Furthermore, unless
exact ground-state degeneracies occur, as for the Ising model with bimodal
$P(J_{ij})$ \cite{bray:87a}, $\theta_s$ is the only non-trivial exponent, while the
critical exponent $\eta$ is simply $2-D$ when $\theta_s<0$ \cite{bray:87a}.
Consequently, 2D models offer a crucial test bench for our understanding of spin
glasses at low temperatures.

Twenty years of research since the original DWRG work of Morris {\it et al.\/}\
\cite{morris:86a} have not been able to settle a number of persistent controversies
concerning the ground-state properties of the 2D {\em XY\/} spin glass.  Firstly, it
has been suggested that the ground state may possess non-trivial extensive
degeneracies when $P(J_{ij})$ is a discrete bimodal distribution
\cite{jain:86a,ray:92a,gingras:92b}. Secondly, it was realized early
\cite{toulouse:79a} that the rotational symmetry of the {\em XY\/} spin glass is
accompanied by a ${\mathbb Z}_2$ symmetry originating from the difference between
proper and improper O($n$) rotations \cite{villain:77a}. It has been suggested that
the resulting Ising-like {\em chirality\/} variables may decouple from the rotational
degrees-of-freedom, leading to different critical behavior for the spin and chiral
variables \cite{kawamura:87a}. For $D=2$, where $T_g=0$, this would entail distinct
spin and chiral stiffness exponents, $\theta_s=-1/\nu_s$ and $\theta_c=-1/\nu_c$,
respectively. Finally, and most noteworthy, previous Monte Carlo (MC) \cite{jain:86a}
and DWRG studies \cite{morris:86a,maucourt:98a,kosterlitz:99a} have yielded rather
inconsistent values for $\theta_s$. This might be partly explained by the difficulty
in obtaining ground state configurations of the model. Parallel alignment of the
spins to their local molecular fields $\bm{h}_i=\sum_j J_{ij}\bm{S}_j$ is a necessary
condition for metastability of the system \eqref{eq:EA_model}.  However, due to the
broad spectrum of an exponential number of metastable states, the resulting commonly
used \cite{morris:86a,maucourt:98a} iterative spin quench algorithm \cite{walker:80a}
almost never yields a ground state configuration.  Additionally, experience with the
simpler 2D Ising case shows that finite-size corrections as well as the dependence on
the chosen pair of boundary conditions are generically large
\cite{hartmann:02a,amoruso:03a}. Hence it seems likely that the observed
inconsistencies for the {\em XY\/} model to date are due to system-size restrictions,
improper finite-size scaling analyses, and limitations in probing the true ground
state behavior.

\begin{figure}[tb]
  \centering
  \includegraphics[clip=true,keepaspectratio=true,width=7cm]{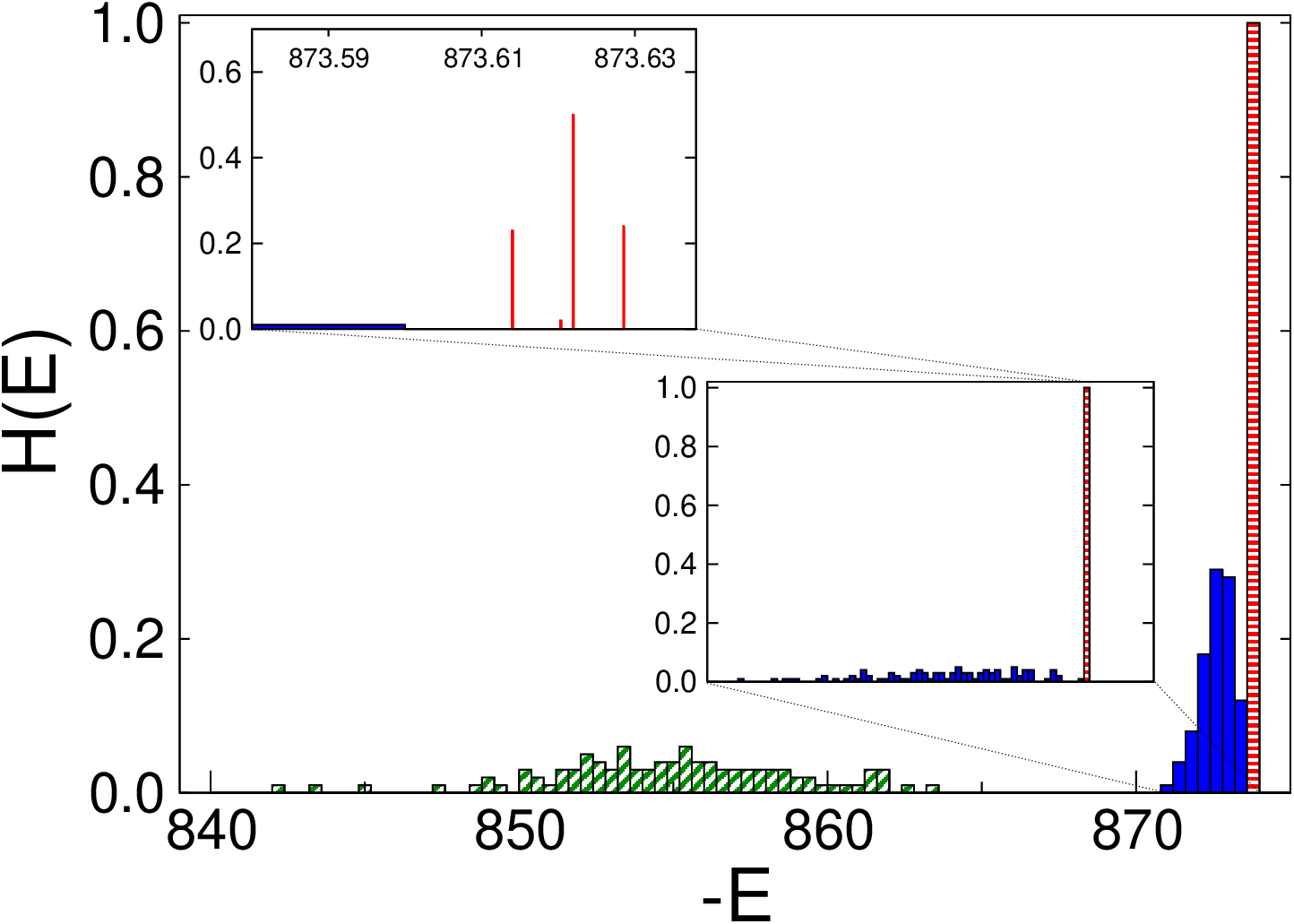}
  \caption
  {Histogram $H(E)$ of minimum energies obtained from repeated runs using the
    spin-quench method (green, diagonally hatched bars), simulated annealing (blue,
    solid bars) and the genetic embedded-matching technique (red, horizontally
    hatched bars) for a {\em single\/} disorder realization $\{J_{ij}\}$ of
    system-size $24\times 24$. The insets show blow-ups of the region around the true
    ground state.  The genetic matching was run here with a small population of
    ${\cal N}_0=64$ replica; for ${\cal N}_0=256$, all runs converge to the rightmost
    bar of the top inset.}
  \label{fig:performance}
\end{figure}
 
Toulouse noted that the ground state problem for the 2D Ising spin glass on a planar
graph can be transformed to a minimization problem for the total length of energy
strings on the dual lattice, connecting pairs of {\em frustrated plaquettes\/}, i.e.,
graph faces containing an odd number of negative bonds
\cite{toulouse:77a,hartmann:book}. It was later realized \cite{bieche:80a} that this
constitutes a {\em minimum-weight perfect matching problem\/}, which is well known in
graph theory and can be solved in polynomial time, such that the ground state of
large 2D Ising spin glass systems can be found exactly. In contrast, the {\em XY\/}
model ground state problem is seemingly not polynomial. However, in this paper we
propose that a partial solution can be found in polynomial time by an embedding of
Ising variables into the continuous spins, allowing us to obtain new results
addressing the controversies alluded to above. This embedding is achieved by choosing
a random direction $\bm{r}$ in spin space to decompose the spins as $\bm{S}_i =
\bm{S}_i^\parallel + \bm{S}_i^\perp = (\bm{S}_i\cdot\bm{r})\bm{r} + \bm{S}_i^\perp$.
A reflection $R_i(\bm{r})$ of $\bm{S}_i$ along the plane defined by $\bm{r}$ maps
$\bm{S}_i^\parallel \rightarrow -\bm{S}_i^\parallel$ and $\bm{S}_i^\perp \rightarrow
\bm{S}_i^\perp$. Hence, with respect to these local reflections the Hamiltonian
\eqref{eq:EA_model} decomposes as ${\cal H} = {\cal H}^{r,\parallel} + {\cal
  H}^{r,\perp}$ with ${\cal H}^{r,\parallel} = -\sum_{\langle
  i,j\rangle}\tilde{J^r_{ij}}\,\epsilon_i^r \epsilon_j^r$, and
\begin{equation}
  \label{eq:embedd}
  \tilde{J_{ij}^r} = J_{ij}|\bm{S}_i\cdot\bm{r}||\bm{S}_j\cdot\bm{r}|,\;\;\;
  \epsilon_i^r = \mathrm{sign}(\bm{S}_i\cdot\bm{r}).
\end{equation}
Thus, since the $R_i(\bm{r})$ merely induce an inversion $\epsilon_i^r \rightarrow
-\epsilon_i^r$, the O($n$) model Hamiltonian \eqref{eq:EA_model} is formally
identified with that of an Ising model, if spin changes are restricted to the
reflections $R_i(\bm{r})$. One can then proceed as follows: decompose the O($n$)
spins $\bm{S}_i$ with respect to $\bm{r}$ and find the corresponding Ising ground
state using the matching technique \cite{bieche:80a}.  This corresponds to a
reflection of some of the $\bm{S}_i$ and thus a new valid O($n$) model configuration.
With ${\cal H}^{r,\perp}$ being invariant, this embedded ground state search
decreases the total energy of \eqref{eq:EA_model} or leaves it constant. The full
O($n$) symmetry can then be statistically recovered by sequential minimizations for a
series $\bm{r}_1,\bm{r}_2,\bm{r}_3,\ldots$ of random directions. We call this
procedure ``embedded matching''. If \eqref{eq:EA_model} is in a ground state, all the
embedded Ising systems must be in (one of) their respective ground state(s) as well.
However, stationarity of the process of successive embedded Ising-like matching
minimization steps does not guarantee global minimum energy for the system
\eqref{eq:EA_model} \cite{weigel:prep}.  Thus, the corresponding artificial dynamics
of exhibits metastability, however with far less metastable states than the local
spin-quench method \cite{weigel:prep}. For further improvement, and to find true
ground states with high reliability, the embedded matching procedure is inserted as a
minimization step into a specially tailored genetic algorithm \cite{weigel:prep}.
Generally speaking, in a genetic algorithm, a population of ${\cal N}_0$ candidate
ground-state configurations is being iteratively optimized by mixing or ``crossing
over'' the ``genetic material'' of different candidate ground states and eliminating
the less well adapted instances \cite{pal:96a}.  To achieve reasonable performance,
this crossover operation has to be chosen appropriately.  Specifically, we are guided
by the direct (visual) inspection of the spin configurations from different
metastable states obtained by the embedded matching technique. There, due to the
local spin rigidity, the predominant differences consist of (proper or improper)
O($n$) rotations of rigid domains. Hence, to preserve the high level of optimization
already obtained {\em inside\/} of domains at intermediate stages of the evolution,
new offspring configurations are produced by randomly exchanging these (automatically
determined) domains instead of single spins between the parent replica.  Full details
of the algorithm will be presented elsewhere \cite{weigel:prep} (for a related method
for the Ising case see Ref.~\cite{hartmann:96a}). This ``genetic embedded-matching''
(GEM) approach works very reliably already for small ${\cal N}_0$ as shown in Fig.\
\ref{fig:performance}. There, we compare the histograms of energies of metastable
states found from statistically independent runs for the same $\pm J$ disorder
configuration $\{J_{ij}\}$ of a $24\times 24$ system using either the simple
spin-quench approach \cite{walker:80a}, the simulated annealing method, or the GEM
technique.  The first two methods give broad distributions of energies, whereas on
this scale, the GEM always seems to yield the same energy, which is clearly below the
range of energies regularly found by the other approaches.  Only on examining the
histograms at much higher resolution, do the GEM data get resolved into a small
series of sharp peaks, corresponding to different energy levels, cf.\ the upper inset
of Fig.\ \ref{fig:performance}. On increasing ${\cal N}_0$ from ${\cal N}_0=64$,
chosen for the runs in Fig.\ \ref{fig:performance}, to ${\cal N}_0=256$, the peaks
displayed in the inset all collapse onto the peak of lowest energy on the right,
corresponding to the true ground state.

As a first result we find, perhaps surprisingly, that the ground states thus obtained
for a given bimodal disorder realization are not only identical in energy up to
machine precision (15 digits) between statistically independent runs, but the final
optimized spin configurations themselves are trivially related to each other by {\em
  global\/} O($n$) transformations \cite{weigel:prep}. In other words, the ground
states are unique and, in contrast to the bimodal Ising model \cite{kawashima:03a},
no accidental degeneracies occur. Hence, after averaging over disorder, the ground
state is ordered and the spin correlation function is constant, implying $\eta=0$
\cite{bray:87a}. This is in contrast to indications by MC simulations
\cite{jain:86a,ray:92a} and Migdal-Kadanoff calculations \cite{gingras:92b}, which
presumably did not probe the true ground state behavior.

\begin{figure}[tb]
  \centering
  \includegraphics[clip=true,keepaspectratio=true,width=7cm]{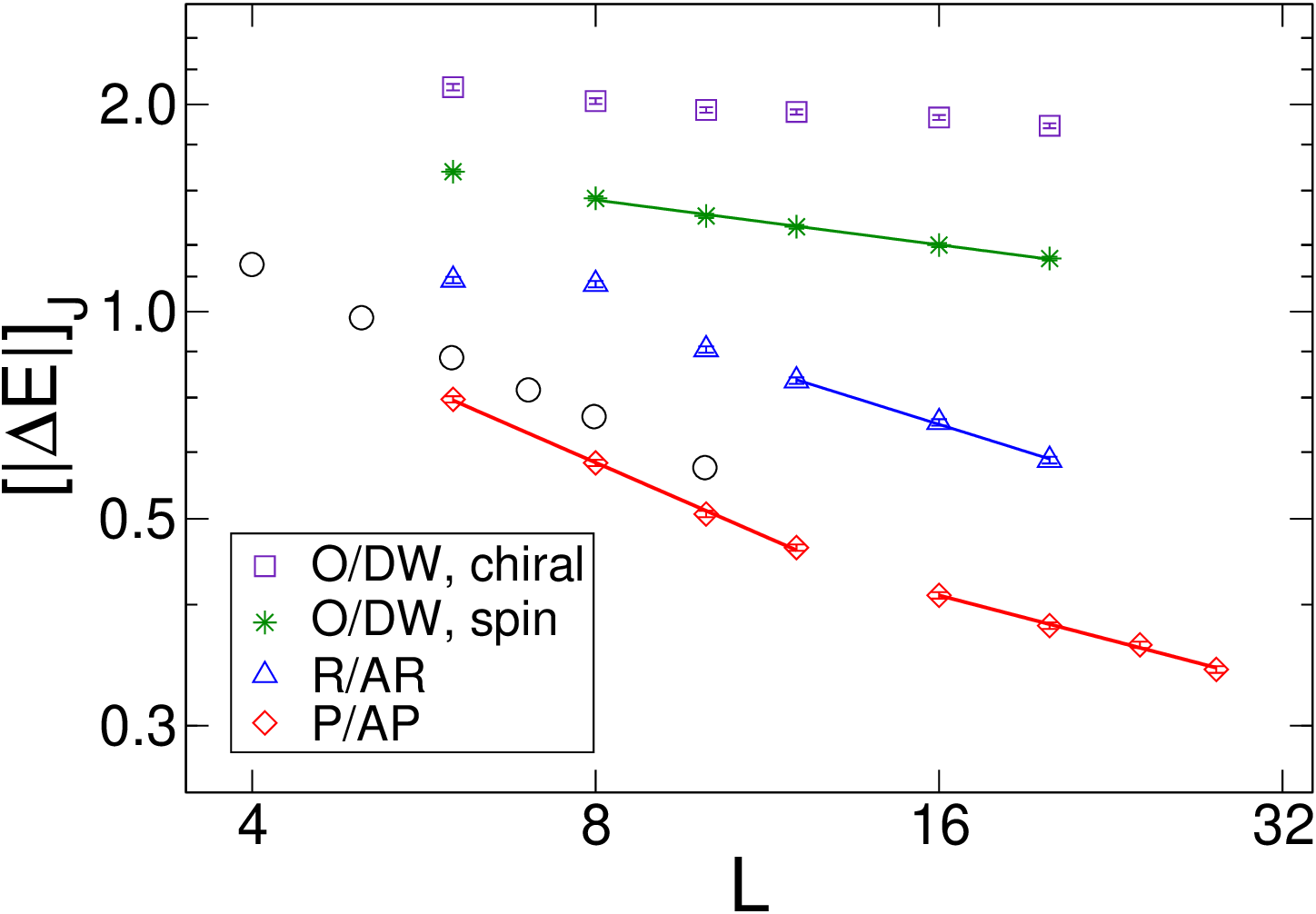}
  \caption
  {Log-log plot of the average domain-wall energies $[|\Delta E|]_J$ for three sets
    of boundary conditions on square lattices as a function of system size $L$. The
    lines are fits of the form $[|\Delta E|]_J\sim L^\theta$ to the data. The black
    circles show the ``random twist'' result of Ref.\ \cite{kosterlitz:99a} for
    comparison. Some data sets have been shifted vertically for better distinction.}
  \label{fig:dw_energy}
\end{figure}

Strong dependence of the domain-wall scaling on the choice of BCs has been observed
for the Ising spin glass \cite{carter:02a,hartmann:02a}. Further complications arise
for the {\em XY\/} case due to the simultaneous presence of continuous (spin) and
discrete (chiral) symmetries: for both periodic (P) and antiperiodic (AP) BCs, domain
walls might be forced into the system due to the periodicity, such that the P/AP
energy difference does not directly capture the energy of single walls.  In
Ref.~\cite{kosterlitz:99a}, it was attempted to alleviate this problem by introducing
a twist along the boundary, which is included in the optimization process to yield
``optimum twist'' BCs. Nevertheless, additional {\em chiral\/} domain-walls might
still occur in the measurements of spin domain-walls.  In fact it has been found that
for (quasi) one-dimensional {\em XY\/} systems {\em both\/} P/AP and reflective BCs
asymptotically probe the chiral excitations \cite{thill:95a}.  These problems,
resulting from a periodic constraint, can be circumvented by applying open and
``domain-wall'' (O/DW) BCs. Here, one ensures the insertion of single domain walls by
comparing the ground state of a system with open BCs to one where the relative
orientations of spins linked across the boundary are either tilted by an angle $\pi$
for spin domain walls or reflected along an arbitrary but fixed axis for chiral
domain walls by the introduction of very strong bonds
\cite{hartmann:01a,weigel:prep}. In addition, and for comparison, we consider P/AP
and random-antirandom (R/AR) BCs as well, the latter fixing the boundary spins in
random relative orientations for one ground state computation (R) and in relatively
$\pi$-rotated orientations for AR BCs. In all cases, the edges with unaltered BCs are
left open. Ground states were computed for systems of up to $28\times 28$ spins,
using 5000 disorder realizations with $J_{ij}=\pm J$ at equal proportions. Figure
\ref{fig:dw_energy} shows the results for the three sets of BCs together with fits of
the asymptotically expected form $[|\Delta E|]_J\sim L^\theta$ to the data, where
$[\cdot]_J$ denotes the average over disorder. The results for P/AP BCs show a
pronounced crossover from $\theta=-0.724(21)$ for $L \le 12$ to $\theta=-0.433(26)$
for $L\ge 16$, the first value being compatible with the ``random twist'' data of
Ref.~\cite{kosterlitz:99a} drawn for comparison ($\theta_s = -0.76$), which are
representative of previous results for P/AP BCs and small system sizes
\cite{maucourt:98a}. On the other hand, $\theta=-0.433(26)$ is closer to the
``optimum twist'' result of Ref.~\cite{kosterlitz:99a}, designed to alleviate the
problem of trapped domain walls.  Note that the apparent crossover length is
compatible with the length below which no metastability occurs and the system behaves
like a spherical spin glass \cite{morris:86a,lee:05a}. The other BCs yield less
negative values already for smaller system sizes, resulting in $\theta_s =
-0.519(30)$ for the R/AR combination and $\theta_s = -0.207(12)$ for the O/DW BCs.
The scaling of the chiral domain-wall energies from O/DW boundaries yields an only
slightly negative value $\theta_c = -0.090(23)$.

\begin{figure}[tb]
  \centering
  \includegraphics[clip=true,keepaspectratio=true,width=7cm]{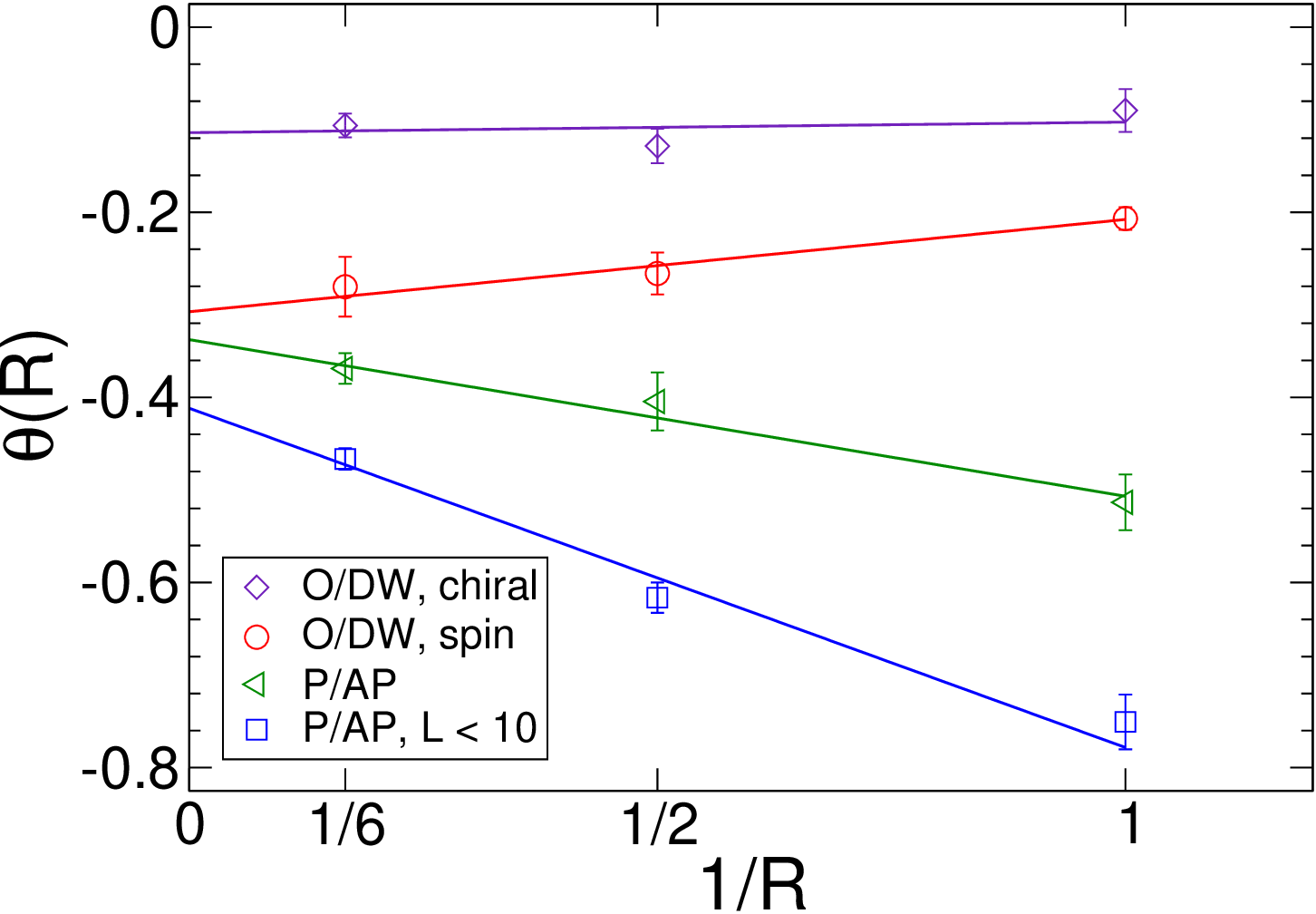}
  \caption
  {Aspect-ratio scaling of the stiffness exponents $\theta_s$ and $\theta_c$ for
    aspect ratios $R=1$, $2$ and $6$ as a function of $1/R$. The bottom data set
    corresponds to fits for the fixed $R$ data restricted to $L\le 10$ (see text).
    The solid lines show fits of the functional form $\theta(R) =
    \theta(R=\infty)+A_R/R$ to the data.}
  \label{fig:ASR_scaling}
\end{figure}

The above usage of multiple pairs of BCs reveals the presence of pronounced
finite-size corrections, even for the already larger system sizes considered here
compared to previous studies \cite{morris:86a,maucourt:98a,kosterlitz:99a}. Part of
these corrections are due to irrelevant scaling fields and sub-leading analytical
terms, giving rise to the general form $[|\Delta E|]_J(L) =
AL^\theta+BL^{-\omega}+C/L+D/L^2+\cdots$.  For a proper resolution of these
contributions, much larger system sizes, out of the reach of current numerical
methods, would be necessary. Thus, we have to restrict ourselves here to a successive
omission of data points from the small-$L$ side to extrapolate towards
$L\rightarrow\infty$. Additional corrections, however, result from the dependence on
the considered pair of BCs. For the Ising system, it has been argued that such
corrections might be suppressed by considering $L\times M$ systems (the change of BCs
happening along the edges of length $L$) with aspect ratios $R\equiv M/L\ne 1$
\cite{carter:02a}.  Neglecting for the time being the corrections listed above, the
asymptotic scaling of defect energies should then follow the form $[|\Delta
E|]_J(L,M) = L^\theta F(R)$ with some scaling function $F$. In general, $F(R)$
depends on the BCs applied \cite{carter:02a}. However, there is no dependence on BCs
for one-dimensional systems \cite{bray:87a,thill:95a}, such that $F(R)$ is
independent of BCs in the limit $R\rightarrow\infty$ and the corresponding
corrections should disappear as more and more elongated systems are being considered.
To investigate this, we determined the ground states of 5000 disorder replica and
$L=4,6,\ldots,16$ for $R=2$ and $L=3,4,\ldots,9$ for $R=6$ in addition to the data
for the square systems ($R\!=\!1$) for the different sets of BCs. Figure
\ref{fig:ASR_scaling} shows the estimated stiffness exponents as a function of $R$
for P/AP and O/DW BCs together with fits of the functional form $\theta(R) =
\theta(R\!=\!\infty)+A_R/R$ to the data, which is inspired by the results for the
Ising case \cite{carter:02a,hartmann:02a}. The scaling corrections at {\em fixed\/}
$R$ listed above are taken into account by including only the largest lattice sizes
in the fits of $[|\Delta E|]_J\sim L^\theta$. For comparison, the bottom dataset of
Fig.\ \ref{fig:ASR_scaling} shows the results from including a fixed range of sizes
$L\le 10$ for each aspect ratio $R$, thus admixing the two correction effects. The
fits result in consistent asymptotic estimates of the spin stiffness exponent of
$\theta_s(R\!=\!\infty)=-0.338(20)$ from P/AP BCs and of
$\theta_s(R\!=\!\infty)=-0.308(30)$ from O/DW BCs, indicating that the asymptotic
regime is indeed being probed. The chiral exponent $\theta_c$, on the other hand,
depends only weakly on $R$, and the asymptotic estimate
$\theta_c(R=\infty)=-0.114(16)$ is clearly different from $\theta_s$.

In conclusion, we have developed a novel quasi-exact algorithm to determine the
ground state of 2D O($n$) spin glasses.  Considering for specificity the 2D {\em
  XY\/} spin glass model with bimodal distribution of random exchange couplings
$J_{ij}$, we have shown from computations for relatively large systems sizes that, as
argued in Ref.~\cite{kosterlitz:99a}, defect-wall calculations from P/AP BCs indeed
suffer from large finite-size corrections due to the periodic constraint.  Using
aspect-ratio scaling, however, they are found to asymptotically yield the same
scaling behavior as the less ambiguous O/DW BCs showing less pronounced corrections,
and we quote the latter result as our final estimate, $\theta_s=-0.308(30)$. This
might be compared with $\theta_s \approx -0.28$ for the 2D Ising case with {\em
  Gaussian} coupling distribution \cite{hartmann:02a,hartmann:01a}. The chiral
exponent is found to be $\theta_c = -0.114(16)$, clearly different from the spin
exponent $\theta_s$, indicating spin-chirality decoupling, and close to the value
$\theta_s=0$ found for the bimodal Ising spin glass \cite{hartmann:01a}.  Yet, no
exact degeneracies as occurring in the latter case are found here. It would be very
interesting to see whether the {\em XY\/} spin glass with Gaussian couplings shows a
different behavior than the $\pm J$ case considered here.  The 2D Heisenberg spin
glass is another exciting problem to explore.

We thank A.\ K.\ Hartmann for useful discussions and a critical reading of the
manuscript. Support for this work was provided by the NSERC of Canada, the Canada
Research Chair Program (Tier I) (M.G), the Canada Foundation for Innovation, the
Ontario Innovation Trust, and the Canadian Institute for Advanced Research.


\end{document}